# That's Not Me! Designing Fictitious Profiles to Answer Security Questions

**Nicholas Micallef**

UNSW Canberra Cyber

School of Engineering and Information Technology

University of New South Wales

Canberra, Australia

n.micallef@adfa.edu.au

**Nalin Asanka Gamagedara Arachchilage**

Optus La Trobe Cyber Security Research Hub

Department of Computer Science and IT

School of Engineering and Mathematical Sciences

La Trobe University

Victoria 3086 Australia

n.arachchilage@latrobe.edu.au

## Abstract

Although security questions are still widely adopted, they still have several limitations. Previous research found that using system-generated information to answer security questions could be more secure than users' own answers. However, using system-generated information has usability limitations. To improve usability, previous research proposed the design of system-generated fictitious profiles. The information from these profiles would be used to answer security questions. However, no research has studied the elements that could influence the design of fictitious profiles or systems that use them to answer security questions. To address this research gap, we conducted an empirical investigation through 20 structured interviews. Our main findings revealed that to improve the design of fictitious profiles, users should be given the option to configure the profiles to make them relatable, interesting and memorable. We also found that the security questions currently provided by websites would need to be enhanced to cater for fictitious profiles.

## Author Keywords

Usable security; Security questions; Fictitious profiles.

## ACM Classification Keywords

H.5.2. Information interfaces and presentation: User Interfaces – User-centered design.

## Fictitious Profiles Design

We defined these fictitious profiles (see Figure 1) ourselves, to mitigate some of the limitations that the current implementation of security questions has. For instance, to minimize the profiles vulnerability to dictionary and guessing attacks, when selecting the profiles, we provided a good mix of low, medium, high entropy (i.e. measure of how predictable the attribute value is[1]) attributes [5] and verified that few attributes have a limited answer space. Also, since these profiles were system-generated, they are already less vulnerable to social engineering attacks.

We started by first identifying the most popular security questions categories (i.e. names, favourites and places) that are used on websites [5,12]. We then added numeric attributes (e.g. finance) because some people are better at remembering numbers, rather than text [11]. The characteristics categories were included to make the fictitious profiles look more real, as proposed in [9]. Similar to [9], we used Fake Name Generator[2] to define the content of the profiles. A male and a female profile were selected so that participants could be provided with the two most common genders (we will include more genders in future studies). The age of the profiles was randomly selected by Fake Name Generator.

**Tayla Dobbie (Female)**
Birthday: August 2, 1974
Age: 42 years old
Tropical zodiac: Leo
BASIC INFO
Mother's maiden name: Kinnear
Father's middle name: Ihssan
Best Friend: Gweneth
Phone: 702-214-1334
Vehicle registration number: 88 8048
FINANCE
Visa: 4716 2953 1995 0309
Expires: 2/2019
CVV2: 341
PLACES
High School Street address: 3822 Ottis Street St Louis, OK 74854
College city name: Philadelphia, PA 19108
First Occupation: Bookkeeper
Address of First Occupation: 3668 Melm Street Providence, RI 02903
PHYSICAL CHARACTERISTICS
Height: 5' 6" (170 centimeters)
Weight: 127.2 pounds (57.8 kilograms)
CHARACTERISTICS
Main Skills: Intuition
Weaknesses: Introvert
FAVOURITES
Pets: Nigel (cat), Hazel (gold fish)
Hobbies: Dietitian
Food: Chicken roast

Figure 1: Fictitious profile (Female)

## Introduction

Due to various limitations [4], some online services (e.g. Facebook and Google) started moving away from using security questions and started using text-based and email-based mechanisms to recover forgotten passwords [4]. However, security questions are still widely adopted [2], and research is still being conducted to mitigate their limitations [15].

The main limitations of security questions are that they are vulnerable to: (1) social engineering attacks [5,17] (e.g. through social networks [12]); (2) dictionary/guessing attacks when answers have a limited answers space [5]; and (3) insider attacks [13], since partners could guess 20% of answers [13]. These limitations could be mitigated by using system-generated information to answer security questions, because they were found to be more secure than users' own answers [1,14]. However, using system-generated information has usability limitations (mainly memorability) [1,14]. To improve the usability of system-generated information, Micallef and Just [9] proposed the design of system-generated information in the form of fictitious profiles. However, no research has investigated the elements that could influence the design of fictitious profiles or systems that use them to answer security questions, with the purpose of improving their usability. Hence, to address this research gap, we conducted 20 structured interviews. We provided participants with 2 fictitious profiles (see Figure 1 shown female profile) and explained how these profiles would be used to answer security questions. In these interviews we asked participants about: (1) the elements that would affect the selection of a fictitious profile; (2) the attributes that they would prefer a fictitious profile to have (or not have); (3) the required level of configurability and availability of the fictitious profiles; and (4) whether they would consider using fictitious profiles to answer security questions.

The main contribution of this work to the usable security field is a set of recommendations that would improve the design of systems that would generate fictitious profiles for answering security questions and systems that use security questions to recover passwords.

## Methodology

We conducted structured interviews with 20 participants, to investigate the elements that could influence the design of fictitious profiles and systems that would use them to answer security questions [18]. University's ethic approval was obtained before starting the interviews.

We recruited participants through word of mouth and social connections. The 20 participants were 5 females and 15 males (mean age=30, (22-45) and med=28). Most participants (15) were post-grads and the rest (5) were employed full-time. Everyone was experienced and confident with using security questions.

Interviews were conducted by one researcher in different places to satisfy the individual needs of the participants (e.g. in the meeting room next to the participant's office). All interviews were conducted in-person. We used a structured interview protocol to understand how users' feedback would affect the design of systems that generate fictitious profiles for security questions and security questions mechanisms that would provide questions based on these profiles.

In the beginning, participants were briefed about the study, were asked to read an information sheet and sign a consent form. Participants also completed a pre-study questionnaire to collect demographic information together with details about their experience with using security questions. Afterwards, we showed participants 2 fictitious profiles – both male and female (see Figure 1 shown Female) and explained that the details of these profiles would be used to answer security questions. Then, we asked questions to understand the elements that would affect the selection of a fictitious profile, the attributes that participants would prefer, the level of configurability and availability that a fictitious profile should have. Finally, we asked participants whether they would consider using fictitious profiles to answer security questions (see box on the left for the exact questions).

---

[1] http://whatis.techtarget.com/definition/password-entropy

[2] http://www.fakenamegenerator.com

> **Questions asked in structured interviews**
>
> **Elements affecting profile selection:**
> Which profile would you use? and Why?
>
> **Attributes selection:**
> On the provided sheet mark the attributes that:
> - the profile should have.
> - the profile should not have.
> - you would add.
>
> **Configurability:**
> Would you like to be able to configure the profile?
> What level of configurability should the profile have?
>
> **Availability:**
> Would you like the profile to be available to you anytime? or do you want to see it just once?
>
> **Potential use:**
> Would you consider using a fictitious profile to answer security questions?
> Explain Why?

*Constant Comparative Method (CCM)*
To identify themes from the collected qualitative information we used an adapted version [3] of the constant comparative method (CCM) approach [6], which has been used in HCI research to analyse qualitative feedback [8]. The interviewer recorded the participants' responses to the interview questions in the form of detailed notes. These notes were later coded by two researchers independently. Both researchers used the created codes to identify common themes from the collected feedback. In most instances, the themes identified/extracted by the two researchers were similar. In the few instances in which there was a disagreement, a third researcher was asked to break the tie. The next section outlines the main themes extracted from these interviews.

## Results

*What elements affect the selection of fictitious profiles?*
To understand the elements that affect the selection of fictitious profiles we asked participants to select one of the provided profiles (male or female - Figure 1: Fictitious profile) and to motivate their choice. The themes extracted from the provided feedback are that the elements that affected the selection of fictitious profile were: (1) relatability/connectedness (e.g. P10 said "I selected the profile because he is male. It is easier for me to associate myself to a male character. Because then I can just compare it to myself."); (2) memorability (e.g. P1 said "Male because there are some things which are memorable"); (3) interesting attributes (e.g. P16 said: "the female one. She seems more interesting. We have a lot of things in common.").

*Are there any preferred attribute categories?*
We asked participants to mark on the provided fictitious profiles the attributes that they would keep, remove and add to the profile (see Figure 2). The main finding from Table 1 is that participants prefer attributes related to basic info (text), characteristics and favourites and would remove numeric attributes (finance). Our results seem to be inconclusive on whether fictitious profiles for security questions attributes should have related to basic info (numbers), places and physical characteristics. Also, our findings reveal that most of our participants would like to add attributes in the favourites and characteristics categories.

*What level of configurability should these profiles have?*
We asked participants about their desired level of configurability in a fictitious profile that would be used to answer security questions. Most participants (14/20) reported that these fictitious profiles should be configurable. However, the qualitative feedback reveals that participants were divided in terms of the desired level of configurability of these profiles. Almost half the participants (11/20) reported that the profile should be highly configurable, meaning that the users should be able to modify specific values (e.g. P15 said "a detailed one because it would be easily memorised"). On the other hand, 9/20 participants reported that the level of configurability should be low, meaning that users should only be able to modify basic attributes, such as age, gender and country (e.g. P6 said "basic stuff, use age range and then generate profile based on that").

*What level of availability should these profiles have?*
Participants were also asked about the level of availability that a fictitious profile should have. Availability is important because it helps understand the security measures that should be used to protect these fictitious profiles. Almost all participants (16/20) reported that they would prefer the fictitious profile to be available all the time, while only 4/20 reported that the profile should have a limited availability. The qualitative analysis of the provided feedback did not reveal any common themes to explain why participants would prefer to access the profile all the time.

*Would users use fictitious profiles and why?*
Finally, we wanted to understand whether participants would consider using fictitious profiles to answer security questions. Almost half the participants (11/20) reported that they would consider using a fictitious profile, while 8/20 reported that they would prefer to use their own answers. Only 1 participant could not makeup his/her mind. The main theme extracted from those participants that reported that they would consider using
a fictitious profile is that it would help them improve the security of their online accounts (e.g. P4 said: "it would be good because no-one would know the answer. That would be more secure than using real answers"). Alternatively, the main theme extracted from those participants that would prefer to use their own answers are that they are not ready to trade-off memorability for an improved security [10] (e.g. P11 said: "because I prefer to answer questions and answers that I'm pretty sure that I can remember").

Figure 2: Example of attributes marked by participants

Table 1: Attributes selection

| | Keep | Remove |
|---|---|---|
| Basic Info (Text) | 17 | 3 |
| Basic Info (Numbers) | 9 | 11 |
| Finance | 3 | 17 |
| Places | 10 | 10 |
| Physical Characteristics | 8 | 12 |
| Characteristics | 16 | 4 |
| Favourites | 20 | 0 |

## Discussion and Recommendations
*Improving the design of fictitious profiles*
Our main findings reveal that our participants prefer fictitious profiles that are highly configurable, to make them relatable, interesting and memorable. Hence, we recommend that designers of systems that would generate fictitious profiles for security questions should implement techniques that prevent users from configuring the fictitious profiles to match their own attributes or to define attributes with a limited answer space, as this would defeat the purpose of having system-generated profiles. This could be achieved by checking that the attributes do not match the users' social networking accounts.

*Compatibility with current security questions*
Our participants seemed to prefer security questions related to characteristics and favourites. Currently, most security questions are about names, places and favourites [5,12]. Hence, if fictitious profiles had to be widely adopted, we recommend that designers of systems that provide security questions as a mechanism to recover forgotten passwords should enhance the security questions that they provide. Otherwise their systems would not cater for those users that would want to use fictitious profiles.

*Availability vs security*
Our findings also reveal that users would prefer fictitious profiles to be available all the time. Thus, more availability of these profiles would increase the possibility that these fictitious profiles could be compromised. Hence, we recommend that system designers should invest a considerable amount of time and effort to implement stronger security measures (e.g. encryption and anonymization techniques) to protect these profiles.

*Improving potential adoption of fictitious profiles*
More than half of our participant would consider using fictitious profiles to answer security questions.
However, the rest of our participants reported that they would prefer to use their own answers due to memorability concerns. This finding seems to indicate that memorability could limit the usability of system-generated information [1,14] even when designed as a fictitious profile. Hence, we recommend that to improve the potential adoption of fictitious profiles when answering security questions, further research needs to investigate techniques that could improve the memorability of these profiles. Very recently [7,16] proposed the use of a serious games to improve the memorability of stronger answers to security questions. However, there is no empirical evidence that validates the effectiveness of these techniques on the long-term.

## Further Research
In our next studies, we will empirically evaluate whether the fictitious profiles designed in this research do actually improve the usability (mainly memorability) of system-generated information when answering security questions. Moreover, since our findings indicate that fictitious profiles seem to have been well received by our participants, we also suggest that further research should be conducted to investigate the design of fictitious profiles for other application areas. For example, to understand how users would design fictitious profiles to anonymise and protect their privacy when registering to online accounts.

# References

1. Mahdi Nasrullah Al-Ameen, Matthew Wright, and Shannon Scielzo. 2015. Towards Making Random Passwords Memorable. *Proceedings of the 33rd Annual ACM Conference on Human Factors in Computing Systems - CHI '15*, ACM Press, 2315–2324. http://doi.org/10.1145/2702123.2702241

2. Yusuf Albayram and Mohammad Maifi Hasan Khan. 2016. Evaluating smartphone-based dynamic security questions for fallback authentication: a field study. *Human-centric Computing and Information Sciences* 6, 1: 16. http://doi.org/10.1186/s13673-016-0072-3

3. Lynne Baillie. 2002. The home workshop: a method for investigating the home. Retrieved Sept 2, 2015 from http://researchrepository.napier.ac.uk/3858/

4. Joseph Bonneau, Elie Bursztein, Ilan Caron, Rob Jackson, and Mike Williamson. 2015. Secrets, Lies, and Account Recovery. *Proceedings of the 24th International Conference on World Wide Web - WWW '15*, ACM Press, 141–150. http://doi.org/10.1145/2736277.2741691

5. Joseph Bonneau, Mike Just, and Greg Matthews. 2010. What's in a Name? Evaluating Statistical Attacks on Personal Knowledge Questions. *International Conference on Financial Cryptography and Data Security*, Springer, Berlin, Heidelberg, 98–113. http://doi.org/10.1007/978-3-642-14577-3_10

6. Barney G Glaser, Anselm L Strauss, and Elizabeth Strutzel. 1968. The Discovery of Grounded Theory; Strategies for Qualitative Research. *Nursing Research* 17, 4.

7. Nicholas Micallef and Nalin Asanka Gamagedara Arachchilage. 2017. A Gamified Approach to Improve Users' Memorability of Fall-back Authentication. *Thirteenth Symposium on Usable Privacy and Security (SOUPS 2017)*, USENIX Association.

8. Nicholas Micallef, Lynne Baillie, and Stephen Uzor. 2016. Time to exercise!: an aide-memoire stroke app for post-stroke arm rehabilitation. *Proceedings of the 18th international conference on Human-computer interaction with mobile devices and services MobileHCI '16*, ACM Press, 112–123. http://doi.org/10.1145/2935334.2935338

9. Nicholas Micallef and Mike Just. 2011. Using Avatars for Improved Authentication with Challenge Questions. *SECURWARE 2011, The Fifth International Conference on Emerging Security Information, Systems and Technologies*, 121–124.

10. Nicholas Micallef, Mike Just, Lynne Baillie, Martin Halvey, and Hilmi Gunes Kayacik. 2015. Why aren't users using protection? Investigating the usability of smartphone locking. *Proceedings of the 17th international conference on Human-computer interaction with mobile devices and services-MobileHCI '15*, ACM Press. http://doi.org/10.1145/2785830.2785835

11. Marisca Milikowski and Jan J. Elshout. 1995. What makes a number easy to remember? *British Journal of Psychology* 86, 4: 537–547. http://doi.org/10.1111/j.2044-8295.1995.tb02571.x

12. *Ariel Rabkin and Ariel. 2008. Personal knowledge questions for fallback authentication:security questions in the era of Facebook. *Proceedings of the 4th symposium on Usable privacy and security - SOUPS '08*, ACM Press, 13. http://doi.org/10.1145/1408664.1408667*

13. Stuart Schechter, A.J. Bernheim Brush, and Serge Egelman. 2009. It's No Secret. Measuring the Security and Reliability of Authentication via "Secret" Questions. *2009 30th IEEE Symposium on Security and Privacy*, IEEE, 375–390. http://doi.org/10.1109/SP.2009.11

14. Shari Trewin, Cal Swart, Larry Koved, Jacquelyn Martino, Kapil Singh, and Shay Ben-David. 2012. Biometric authentication on a mobile device: A Study of User Effort, Error and Task Disruption. *Proceedings of the 28th Annual Computer Security Applications Conference on - ACSAC '12*, ACM Press, 159–168.

15. Peng Zhao, Kaigui Bian, Tong Zhao, et al. 2017. Understanding Smartphone Sensor and App Data for Enhancing the Security of Secret Questions. *IEEE Transactions on Mobile Computing* 16, 2: 552–565. http://doi.org/10.1109/TMC.2016.2546245

16. Micallef, Nicholas, and Nalin Asanka Gamagedara Arachchilage. "Security questions education: exploring gamified features and functionalities." Information & Computer Security 26, no. 3 (2018): 365-378.

17. Arachchilage, Nalin Asanka Gamagedara, Steve Love, and Konstantin Beznosov. "Phishing threat avoidance behaviour: An empirical investigation." Computers in Human Behavior 60 (2016): 185-197.

18. Arachchilage, Gamagedara, and Nalin Asanka. Security awareness of computer users: A game based learning approach. Diss. Brunel University, School of Information Systems, Computing and Mathematics, 2012.